\documentstyle[multicol,aps,prl]{revtex}

\begin{document}

\title{Strange Star Heating Events as a Model for 
Giant Flares of Soft Gamma-ray Repeaters}

\author{V.V. Usov}

\address{Department of Condensed Matter Physics, Weizmann Institute
of Science, Rehovot 76100, Israel}

\maketitle

\begin{abstract}
Two giant flares were observed on 5~March 1979 and 27 August 1998
from the soft $\gamma$-ray repeaters SGR 0526-66 and SGR 1900+14,
respectively. The striking similarity between these remarkable
bursts strongly implies a common nature. We show 
that the light curves of the giant bursts may be easily explained 
in the model where the burst radiation is produced by the bare quark 
surface of a strange star heated, for example, by impact
of a massive comet-like object.

\end{abstract}


\begin{multicols}{2}

I. {\it Introduction}.--
Strange stars are astronomical compact objects which are entirely
made of deconfined quarks. The possible existence of strange 
stars is a direct consequence of the conjecture by Witten \cite{W84}
that strange quark matter (SQM) composed 
of roughly equal numbers of up, down, and strange quarks 
plus a smaller numbers of electrons (to neutralize the electric
charge of the quarks) may be the 
absolute ground state of the strong interaction, i.e., absolutely
stable with respect to $^{56}$Fe. SQM has been studied in many papers
(e.g., Ref. \cite{FJ84}),
and it was shown that, with the uncertainties inherent in a 
nuclear-physics calculation, the existence of stable SQM is plausible.
The bulk properties (size, moment of inertia, etc.) of models 
of strange and neutron stars in the observed mass range $(1< M/M_\odot
<1.8)$ are rather similar, and it is very difficult to discriminate 
between strange and neutron stars \cite{HZS86,G97}. SQM with the density 
of $\sim 5\times 10^{14}$ g~cm$^{-3}$ can exist, by hypothesis,
up to the surface of strange stars \cite{G97,AFO86}. 
Such a bare strange star differs qualitatively from a neutron star 
which has the density at the stellar surface (more exactly at the
stellar photosphere) of about $0.1-1~{\rm g~cm}^{-3}$. This opens 
observational possibilities to distinguish strange stars from neutron
stars, if indeed the formers exist.

Since SQM at the surface of a bare strange star is bound via strong
interaction rather than gravity, such a star is not subject to
the Eddington limit and can radiate at the luminosity greatly 
exceeding $L_{\rm Edd}\simeq 1.3\times 10^{38}
(M/M_\odot)$~ergs~s$^{-1}$ \cite{AFO86}.
Therefore, bare strange stars are reasonable candidates
for soft $\gamma$-ray repeaters (SGRs) that are the sources of 
flares with 
Super-Eddington luminosities, up to $\sim 10^{44}-10^{45}~{\rm ergs~s}
^{-1}$.

There are four known SGRs; three within our Galaxy (SGR~1900+14,
SGR 1806-20, and SGR 1627-41) and one is in the Large Magellanic Cloud
(SGR 0526-66). 
SGRs appear to be associated with radio supernova remnants, indicating
that they are young ($\lesssim 10^4$~yr). SGRs are characterized by their 
recurrent emission of brief ($\sim 0.1$~s), intense ($\sim 10^3-10^4
\,L_{\rm Edd}$) bursts with soft $\gamma$-ray spectra~\cite{K95}.

A remarkable flare was observed by nine satellites 
on 5~March 1979 \cite{M79}. It was the first burst recorded 
from SGR~0525-66. The location of SGR~0525-66 is consistent with a 
supernova remnant (N49) in the Large Magellanic Cloud. Assuming 
a distance of 50~kpc to the supernova remnant N49, the peak luminosity 
of the short ($\sim 0.25$~s) initial pulse was $\sim 1.6\times
10^{45}~{\rm ergs~s}^{-1}$ \cite{FKL96}, seven orders of magnitude in 
excess of the Eddington limit for a solar-mass object. This luminosity 
is about ten times higher than the luminosity of our Galaxy.
After the initial pulse, the source was observed for at least
200~s and pulsated with an 8~s periodicity, which was inferred to be 
the rotational period of SGR~0526-66. Recently (August 27, 1998),
a giant burst was observed from SGR 1900+14 \cite{H99}. This burst 
is nearly a carbon copy of the 5 March 1979 event (see Table~1). 

The model where the source of the 5 March 1979 event is a strange star
has been long ago proposed by Alcock, Farhi, and Olinto \cite{AFO86b}. 
Later, a few other strange star models were developed for SGRs 
\cite{CD98,ZXQ}. However, the light curves expected for bursts in all 
these models were never calculated because the thermal emission from the
bare quark surface of a strange star was poorly known. Recently,
the thermal emission of bare strange stars 
was considered \cite{U98,U00}, and it was shown that 
creation of $e^+e^-$ pairs by the Coulomb barrier at the quark surface
is the main mechanism of thermal emission from the surface of SQM
at the temperature $T_{\rm s} < 5\times 10^{10}$~K. 
Created $e^+e^-$ pairs mostly annihilate in the vicinity of the 
strange star into $\gamma$-rays. In this Letter,
using the results of \cite{U98,U00} we show that the light 
curves of the two giant bursts may be easily explained in the model 
where the burst radiation is produced by the 
bare surfaces of strange stars heated up to $\sim 2\times 10^9$~K 
by impacts of massive comet-like objects.  

II. {\it The model}.-- Imagine that a comet-like object with the mass
$\Delta M\sim 10^{25}$~g falls onto a strange star. We 
assume that the comet matter accretes steadily and spherically.
The total duration of the accretion is $\Delta t \sim 10^2-10^3$~s. 
The accreted matter sinks into the strange star and quarkonizes
\cite{AFO86}. During the accretion, $t<\Delta t$,
the surface layers of the strange star are heated, while their 
thermal radiation is completely suppressed by the falling matter. 
The total thermal energy accumulated in the surface layers at the
moment $t=\Delta t$ is $Q\simeq 0.1\Delta M c^2\sim 10^{45}$~ergs.
When the accretion is finished and the strange star vicinity is
transparent for radiation, some part of the energy $Q$ 
may be emitted from the quark surface and observed as a giant burst.

In our case
the thickness of the surface layer which is heated by accretion 
is very small compared with the stellar radius $R\simeq 10^6$~cm 
(see below), and a plane-parallel approximation may be used. 
We start with the equation of hear transfer that describes
the temperature distribution at the surface layers of
a strange star \cite{I82}:

\begin{equation}
C_q{\partial T\over \partial t}={\partial \over \partial x}
\left(K_c{\partial T\over \partial x}\right) -\varepsilon_\nu\,,
\end{equation}

\noindent
where

\begin{equation}
C_q\simeq 2.5\times 10^{20}(n_b/n_0)^{2/3}T_9\,\,\,
{\rm ergs~cm}^{-3}~{\rm K}^{-1}
\end{equation}

\noindent
is the specific heat for SQM per unit volume,

\begin{equation}
K_c\simeq 6\times 10^{20}\alpha_c^{-1}
(n_b/n_0)^{2/3}\,\,\,
{\rm ergs}~{\rm cm}^{-1}~{\rm s}^{-1}~{\rm K}^{-1}
\end{equation}

\noindent
is the thermal conductivity, 

\begin{equation}
\varepsilon_\nu \simeq 2.2\times 10^{26}\alpha_c 
Y^{1/3}_e(n_b/n_0)T^6_9\,\,\,{\rm ergs}~{\rm cm}^{-3}~{\rm s}^{-1}
\end{equation}

\noindent is the neutrino emissivity,
$n_0\simeq 1.7\times 10^{38}$~cm$^{-3}$ is normal nuclear matter
density, $n_b$ is the baryon number density of SQM,
$\alpha_c=g^2/4\pi$ is the QCD fine structure constant, 
$g$ is the quark-gluon coupling constant, 
$Y_e=n_e/n_b$ is the number of electrons per baryon,
and $T_9$ is the temperature in units of $10^9$~K.

The heat flux due to thermal conductivity is

\begin{equation}
q=-K_cdT/dx\,.
\end{equation}

\noindent
At the stellar surface, the heat flux 
is directed into the strange star and coincides with the energy 
flux of the accreted matter at $0\leq t <\Delta t$,
while at $t\geq \Delta t$
this flux is directed outside and coincides with the energy 
flux in $e^+e^-$ pairs emitted from the SQM surface:

\begin{equation}
q\simeq
\cases{Q/(4\pi R^2\Delta t)
&\quad at $0\leq t < \Delta t$\,,
\cr
\noalign{\vskip3pt}
- \varepsilon_\pm f_\pm
&\quad at
$t\geq \Delta t$\,,
\cr}
\end{equation}

\noindent
where $\varepsilon_\pm\simeq m_ec^2 + kT_{\rm s}$ 
is the mean energy of created $e^+e^-$ pairs,

\begin{equation}
f_\pm\simeq 10^{39.2}\,T_{\rm s,9}
^3\exp \left(-{11.9\over
T_{\rm s,9}}\right)J(\zeta)\,\,\,{\rm cm}^{-2}\,{\rm s}^{-1}
\end{equation}

\noindent
is the flux of pairs from the unit SQM surface,

\begin{equation}
J(\zeta )={1\over 3}{\zeta^3\ln \, (1+2\zeta ^{-1})\over 
(1+0.074\zeta )^3}
+ {\pi^5\over6}{\zeta^4\over (13.9 +\zeta)^4}\,,
\end{equation}

\noindent
and $\zeta\simeq (2\times 10^{10}~{\rm K})/T_{\rm s}$
\cite{U00}.

Eqs. (5)-(8) give a boundary condition on $dT/dx$ at the stellar
surface.  We assume that at the initial moment,
$t=0$, the temperature in the surface layers is constant, 
$T=3\times 10^7$~K. In our model there are two
parameters, $Q$ and $\Delta t$, which describe the 
comet matter accretion onto the strange star.

III. {\it The light curves}.-- The set of Eqs. (1)-(8) was solved 
numerically. 
We assumed the typical values of $\alpha_c=0.1$, $n_b=2n_0$, 
and $Y_e=10^{-4}$. For $Q=9.2\times 10^{44}$~ergs and $\Delta t =370$~s,
Figures~1 and~2 show the luminosity, 
$L_\pm = 4\pi R^2\varepsilon_\pm f_\pm$, of the strange star
in $e^+e^-$ pairs as a function of time $t$ at $t\geq \Delta t$. 
This luminosity is many orders of magnitude higher than

\begin{equation}
L_\pm ^{\rm max}\simeq 4\pi m_{\rm e}c^3R/\sigma_{_{\rm T}}\simeq
10^{36}\,\,{\rm ergs~s}^{-1}\,,
\end{equation}

\noindent
where $\sigma_{_{\rm T}}$ is the Thomson cross-section.
In this case, $e^+e^-$ pairs outflowing from
the stellar surface mostly annihilate in the vicinity 
of the strange star, $r\sim R$, and far from the star,
$r\gg R$, the luminosity in pairs cannot be significantly
more than $L_\pm^{\rm max}$ \cite{GS85}.
Therefore, at $r\gg R$ the luminosity in X-ray and $\gamma$-ray 
photons practically coincides with the calculated value of $L_\pm$, 
$L_\gamma\simeq L_\pm -L_\pm^{\rm max}\simeq L_\pm$. 

The light curve predicted in our model
for $Q=9.2\times 10^{44}$~ergs and $\Delta t=370$~s (see
Figs.~1 and~2) is in good agreement
with the light curve observed for the 5 March 1979 event (see Table~1).
This is the first earnest evidence that SGRs are 
strange stars, not neutron stars as usually assumed. 
It is worth noting that the theoretical light curve shown by Figures~1
and 2 is averaged over 
10~ms that is the highest time resolution of the observations made by 
the Pioneer Venus Orbiter \cite{FKL96}. From Table~1 we can see  
that the light curve of the 27 August 1998 event may be fitted 
fairly well in our model for $Q=5.4 \times 10^{44}$~ergs and 
$\Delta t =280$~s.

The surface layers heated by the accretion radiate in low-energy
($\lesssim 1$~MeV) neutrinos about one per cent of the total thermal
energy $Q$ (see Table~1). The neutrino light curve expected in our 
model for the 5~March 1979 event is shown by Figure~3. 

IV. {\it Discussion}.-- One of the sources of matter that falls onto 
a strange star producing a SGR could be debris formed in collisions 
of planets orbiting the star in nearly coplanar orbits \cite{K94}. 
In this particular model, there appear two typical masses ($\sim 
10^{25}$~g and $\sim 10^{22}$~g) available 
for prompt infall. Accretion of comet-like objects with the first 
typical mass ($\Delta M \sim 10^{25}$g) may result in 
the giant flares of SGRs
as discussed above. The accretion time depends on $\Delta M$ and
the impact parameter $s$. For $\Delta M\sim 10^{25}$~g and
$s$ less than the tidal breakup radius $r_t$ ($\sim 10^{11}$~cm),
this time is somewhere between $\sim l_c/v(l_c)\sim 0.1$~s and 
$\sim r_t/v(r_t)\sim 10^3$~s if the kinematic viscosity is high 
enough, where $l_c\sim 10^8$~cm is the comet radius, and $v(r)\simeq
(GM/r)^{1/2}$ is the velosity at the distance $r$ from 
the strange star of mass $M$ \cite{K94}. The accretion time of 
$\sim 300$~s (see Table~1) is in the allowed range and seems 
reasonable.

Figure~4 shows the distribution of temperature in the surface
layers at the moment $t=\Delta t$ when the accretion is just 
finished and the powerful radiation from the stellar surface 
just starts. This distribution completely determines the 
subsequent radiation from the strange star at $t\geq \Delta t$.
If the surface layers of a bare strange star are heated very fast 
($\lesssim 10^{-3}$~s) to the temperature shown by 
Figure~4 by any other mechanism, for example by decay of
superstrong ($\sim 10^{14}-10^{15}$~G) magnetic fields \cite{U84}, 
the light curve of the subsequent radiation coincides with the light
curve calculated above and shown by Figures~1 and~2. The energy 
released by the magnetic field decay may be communicated to the
surface by stellar pulsations, rather than any other mechanism 
\cite{R80}. The sound-wave crossing time through the strange star
is $\sim 10^{-4}$~s, which is less than the upper limits in the
rise time of the two giant bursts. The superstrong magnetic field
can confine the radiating $e^+e^-$ plasma \cite{R80}. This may be
tested by observations of giant bursts \cite{F00} and the 
existence of superstrong magnetic fields may be verifyed. 

In our model for SGRs, $e^+e^-$ pairs are the main
component of the thermal emission from the stellar surface 
\cite{U98,U00}. In $\sim 10^4$~s after a giant burst, when the surface
luminosity in pairs is $\sim L_\pm ^{\rm max}\sim 10^{36}$~ergs~s$^{-1}$,
the annihilation radiation with the luminosity of $\sim L_\pm^{\rm max}$
escapes from the stellar vicinity more or
less freely, and its spectrum is a very wide ($\Delta E/E\simeq 0.3$)
line of energy $E\simeq 0.5$~MeV. Observations of such a line with
the $\gamma$-ray spectrometer SPI in the forthcoming INTEGRAL mission
can clarify the nature of SGRs.

I thank anonymous referees for many helpful suggestions that
improved the final manuscript.
This work was supported by the Israel Science Foundation of
the Israel Academy of Sciences and Humanities.

\vfill\eject

\end{multicols}

\begin{table}
\caption{Comparison of observational [8] and theoritical
characteristics of the two giant bursts. The accuracy of the
observational characteristics of the burst radiation is not
higher than $\sim 20$\%.}
$$
\begin{array}{ccddccd}
\tableline
\tableline
 &   \rm{SGR}~0526-66  &\,\,\,\,\,\,\,\,\,\,\,\,\,\,\,\rm{SGR}~1900+14 \\
\tableline
\rm {\bf{Giant~outburst}}  & \rm{March}~5,~1979 & \,\,\,\,\,\,\,\,\,\,\,\,
\,\,\,\rm{August}~27,~1998 \\
\tableline
\rm{\bf{Distance}}  & 50~{\rm kpc}&\,\,\,\,\,\,\,\,\,\,\,\,
\,\,\,10~{\rm kpc}\\
\tableline
  &  \rm{observations}\,\,\,\,\,\,\,\,\,\,\, \,\,\,\rm{theory} &
  \,\,\,\,\,\,\,\,\,\,\rm{observations}\,\,\,\,\,\,\,\,\,\,\,\,\,\,\,\, 
   \rm{theory}\\
\tableline
\rm {\bf{Accretion~of~matter}} &  &\\
{\rm Duration}~\Delta t,~{\rm s} &\,\,\,\,\,\,\,\,\,\,\,\, \,\,\,\,\,\,\,\,\,
\,\,\,\,\,\,\,\,\,\,\,\,\,\,\,\,\,\,\,\,\,\,\,\,370 &\,\,\,\,\,\,\,\,\,
\,\,\,\,\,\,\,\,\,\,\,
\,\,\,\,\,\,\,\,\,\,\,\,\,\,\,\,\,\,\,\,\,\,\,\,\,\,\,\,\,\,\,\,\,\,\,
\,\,\,\,\,\,280\\
{\rm Energy~release}~Q,~{\rm ergs} & \,\,\,\,\,\,\,\,\,\,\,\,\,\,\,\,\,\,\,\,\,
\,\,\,\,\, \,\,\,\,\,\,\,\,\,\,\,\,\,\,\,\,\,\,\,\,\,\,\,\,9.2\times 10^{44}
\, & 
\,\,\,\,\,\,\,\,\,\,\,\, \,\,\,\,\,\,\,\,\,\,\,\,
\,\,\,\,\,\,\,\,\,\,\,\,\,\,\,\,\,\,\,\,\,\,\,\,\,
\,\,\,\,\,\,\,\,\,\,\,\,\,\,\,\,\,\,\,5.4\times 10^{44}~~\\
\tableline
\rm {\bf{Initial~pulse}} &      &   \\
\rm {Duration,~s} & \,\,\,\,\,\,\sim 0.25\,\,\,\,\,\,\,\,\,\,\,\,\,\,\,\,
\,\,\,\,\,\sim 0.2&\,\,\,\,\,\,\,\,\,\,\,\,\,\,\,\,\sim 0.35
\,\,\,\,\,\,\,\,\,\,\,\,\,\,\,\,\,\,
\,\,\,\,\,\,\sim 0.3\\
\rm {~~~~~Peak~luminosity,~ergs~s^{-1}~~~~~} & \,\,\,\,\,\,\,\,
\,1.6\times 10^{45}
\,\,\,\,\,\,\,\,\,\,\,\,\,1.4\times 10^{45}  &  
\,\,\,\,\,\,\,\,\,\,\,\,\,\,\gtrsim\,3.7\times 10^{44}
\,\,\,\,\,\,\,\,\,\,\,\,\,\,\,\,\,\,\,\,\, 4\times 10^{44}

\\
\rm {Energy~release,~ergs} & \,\,\,\,\,\,\,\,\,1.3\times 10^{44}\,\,\,
\,\,\,\,\,\,\,\,\,\,\,\,\,\,\,\,10^{44}\,\,\,\,\,\,\,\,\,
& \,\,\,\,\,\,\,\,\,\,\,\,\,\,\gtrsim\,6.8\times 10^{43}\,\,\,\,\,
\,\,\,\,\,\,\,\,\,\,\,\,\,\,\,\,5\times 10^{43}  
  \\
\tableline
\rm {\bf{Tail}} & &\\  
\rm {Exponential~decay,~s} &\,\,\,\,\sim 100\,\,\,\,\,\,\,\,\,\,\,\,\,\,\,\,\,
\,\,\,\sim 100& \,\,\,\,\,\,\,\,\,\,\,\,\,\,\,\,\sim 80\,\,\,\,\,\,\,\,\,\,\,
\,\,\,\,\,\,
\,\,\,\,\,\,\,\,\,
\,\,\,\sim 80\\ 
\rm {Energy~release,~ergs} & \,\,\,\,\,\,\,\,\,\,\,\,\,3\times 10^{44}
\,\,\,\,\,\,\,\,\,\,\,\,\,\,3.3\times 10^{44} & \,\,\,\,\,\,\,\,\,\,\,\,\,\,
\gtrsim\,5.2\times 10^{43}\,
\,\,\,\,\,\,\,\,\,\,\,\,\,\,\,1.2\times 10^{44}\\ 
\tableline
\rm {Total~energy~release\,\,} &     &    \\
{\rm in~radiation,~ergs} & \,\,\,\,\,\,\,\,\,4.3\times 10^{44}
\,\,\,\,\,\,\,\,\,\,\,\,\,4.3\times 10^{44} &  
\,\,\,\,\,\,\,\,\,\,\,\,\,\gtrsim\,1.2\times 10^{44}
\,\,\,\,\,\,\,\,\,\,\,\,\,\,\,\,1.7\times 10^{44} \\ 
\tableline
\rm {Energy~release}  &  &  \\
\rm {in~neutrinos,~ergs} &\,\,\,\,\,\,\,\,\,\,\,\,\,\,\,\,\,\,\,\,\,\,\,
\,\,\,\,\, \,\,\,\,\,\,\,\,\,\,\,\,\,\,\,\,\,\,\,\,1.4\times 10^{43} & 
\,\,\,\,\,\,\,\,\,\,\,\, \,\,\,\,\,\,\,\,\,\,\,\,
\,\,\,\,\,\,\,\,\,\,\,\,\,\,\,\,\,\,\,\,\,\,\,\,\,
\,\,\,\,\,\,\,\,\,\,\,\,\,\,\,\,\,\,\,\,\,\,2.5\times 10^{42}~~\\
\tableline
\tableline
\end{array}
$$
\end{table}


\par\vfill\eject

\noindent
Figure captions

\medskip

\noindent
Fig. 1. The light curve expected in our model for 
$Q=9.2\times 10^{44}$~ergs and $\Delta t=370$~s.

\noindent
Fig. 2. The initial pulse of the light curve shown in Figure~1.

\noindent
Fig. 3. The luminosity in neutrinos as a function of time for
$Q=9.2\times 10^{44}$~ergs and $\Delta t=370$~s.

\noindent
Fig. 4. The distribution of temperature in the surface layers 
at the moment $t=\Delta t =370$~s.

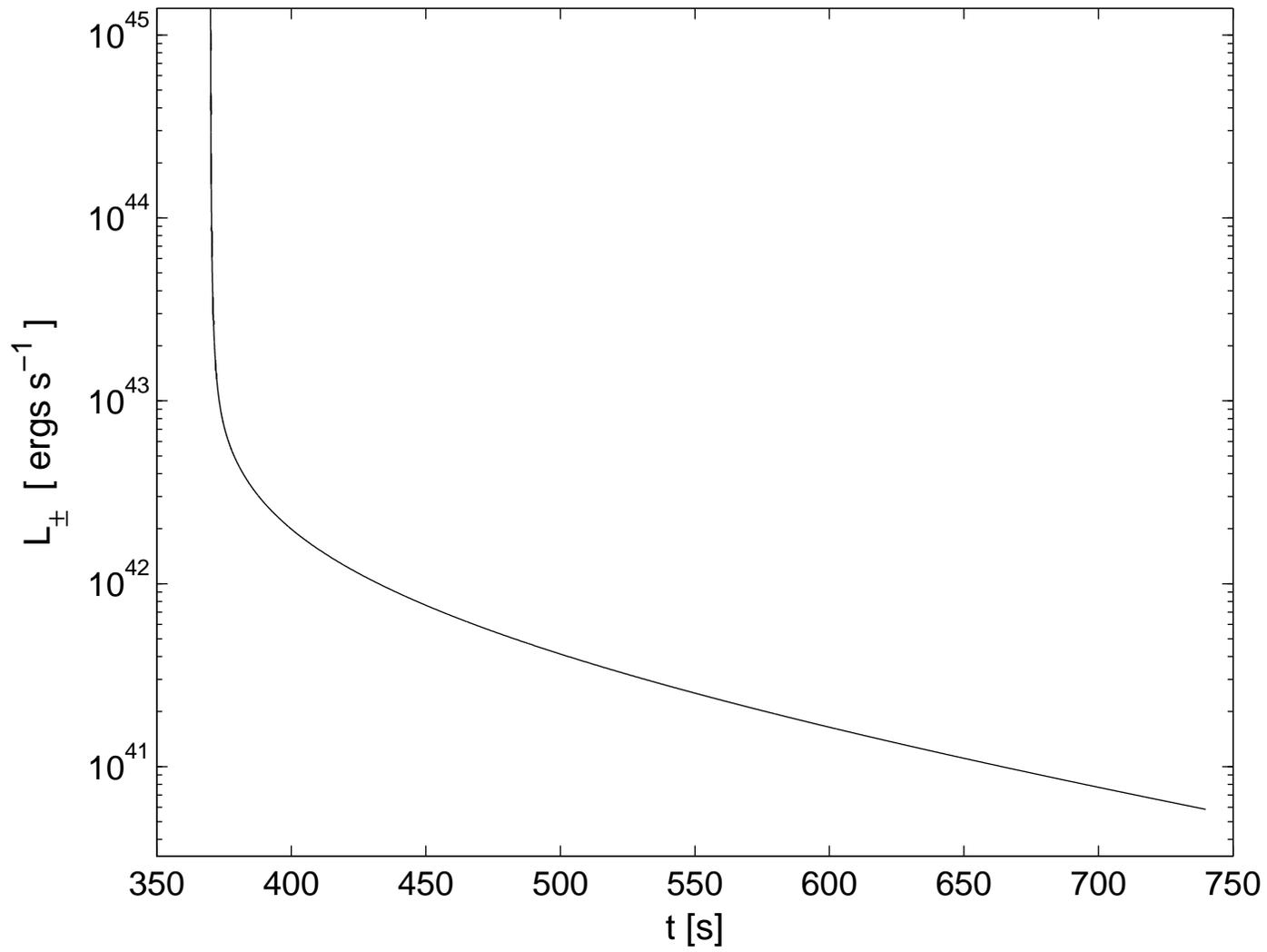
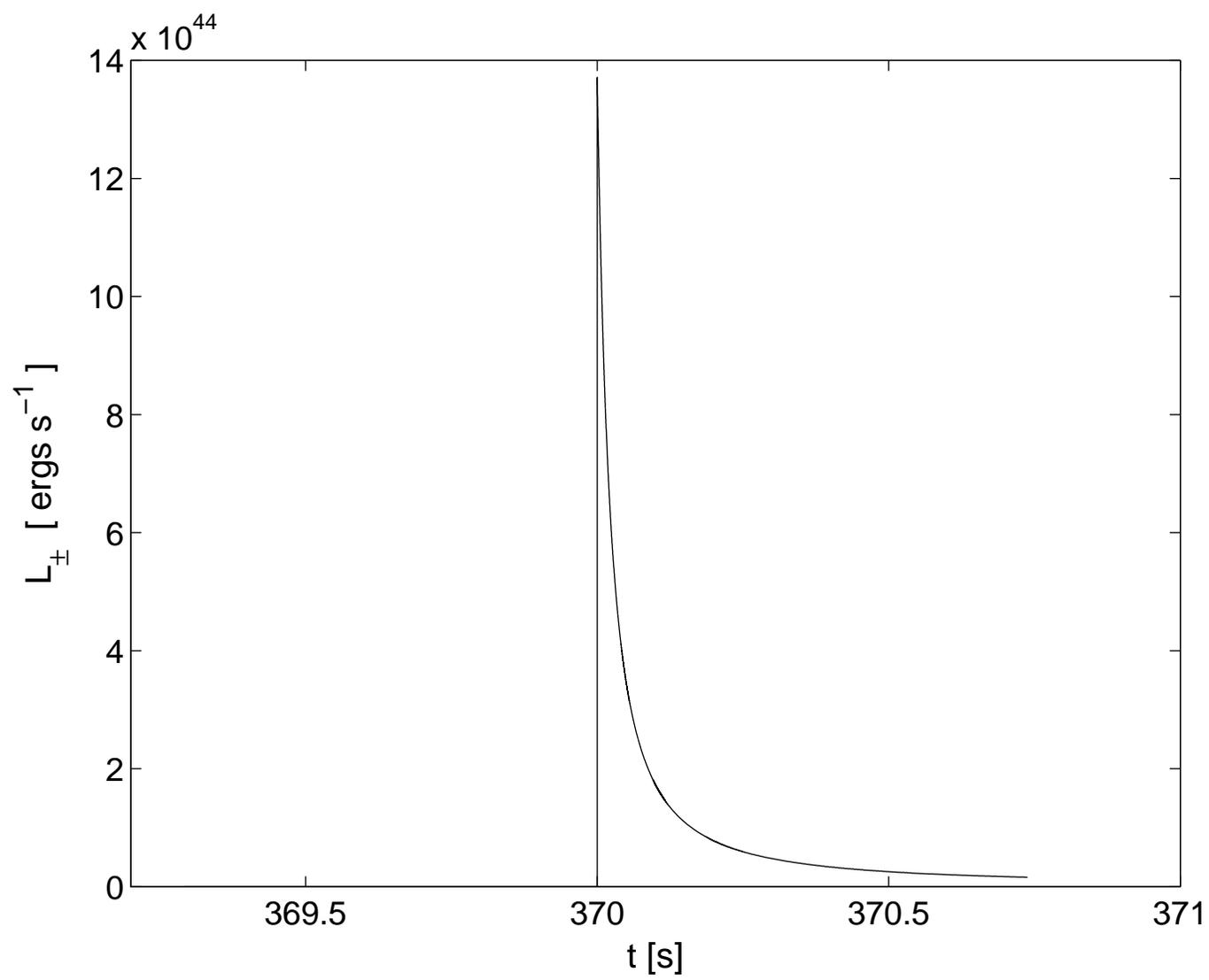
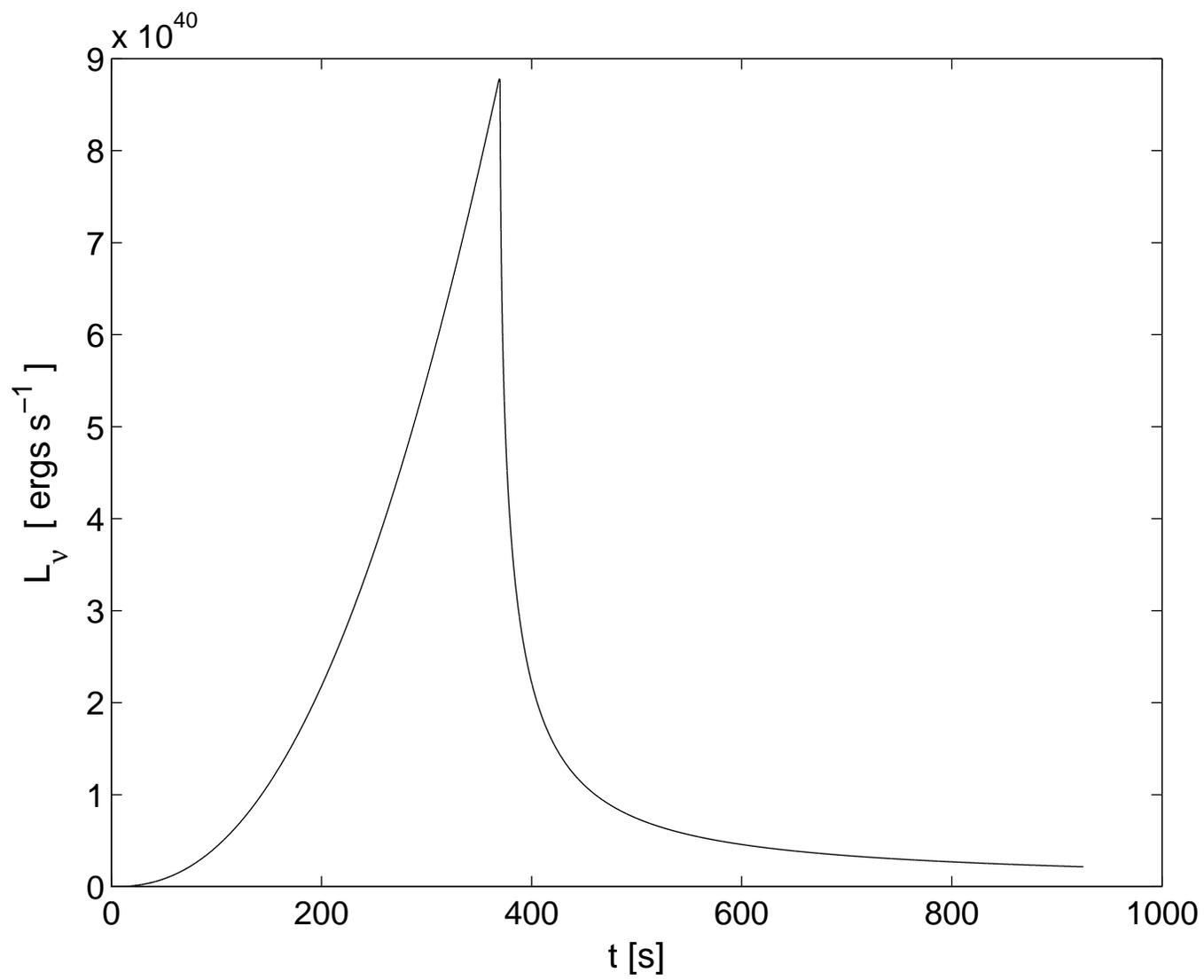
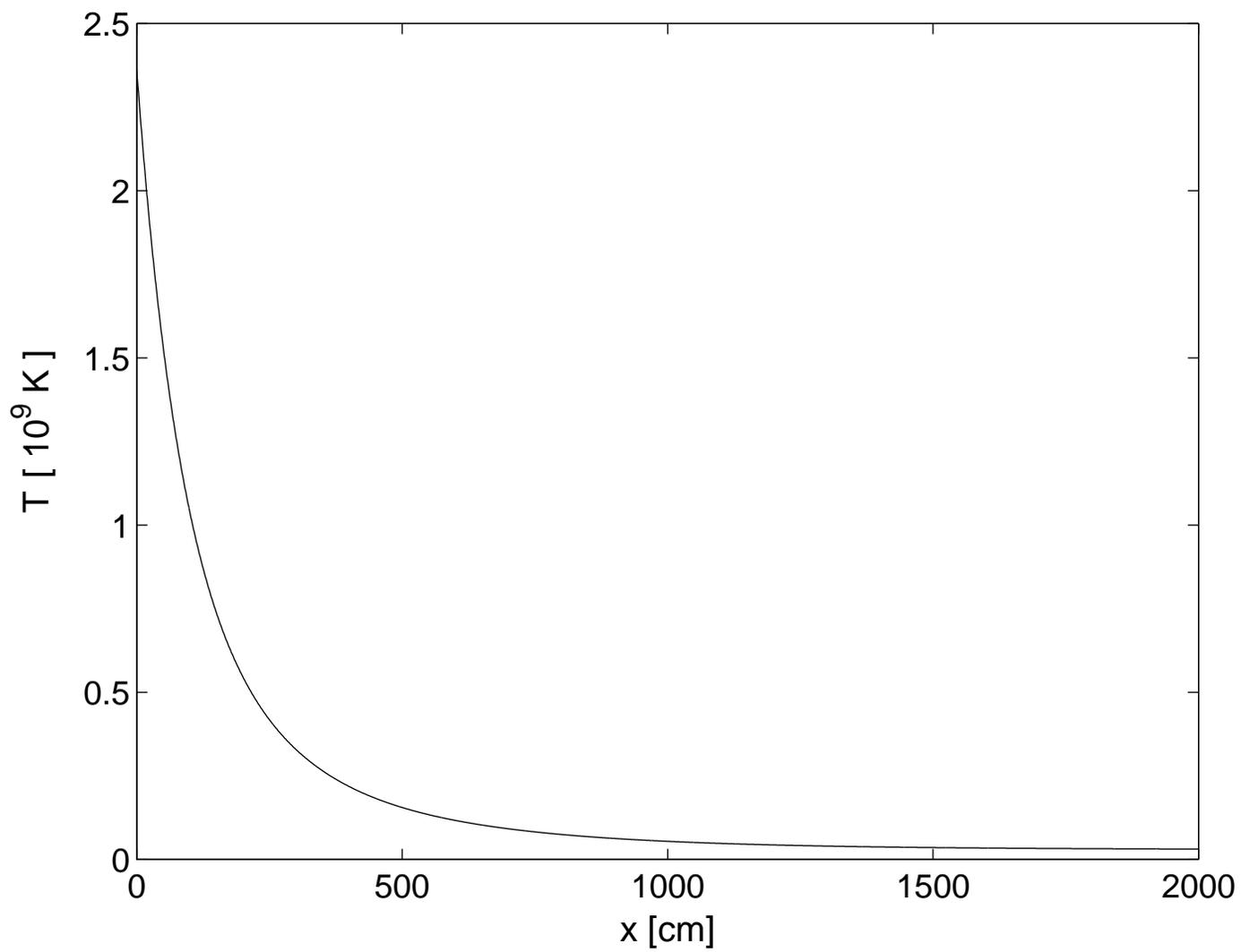
\end{document}